\documentclass[twocolumn,showpacs,preprintnumbers,amsmath,amssymb]{revtex4}
\usepackage{graphicx}% Include figure files
\usepackage{dcolumn}% Align table columns on decimal point
\usepackage{bm}% bold math

\begin{document}

\preprint{APS/123-QED}

\title{Strangelets accelerated by pulsars in galactic cosmic rays}

\author{K.S. Cheng$^1$ and V.V. Usov$^2$}
\affiliation{ $^1$ Department of Physics, The University of Hong
Kong, Hong Kong SAR, People's Republic of China
\\ $^2$Center for Astrophysics, Weizmann Institute, Rehovot
76100, Israel \\}

\date{\today}% It is always \today, today,
             %  but any date may be explicitly specified

\begin{abstract}
It is shown that nuggets of strange quark matter may be extracted
from the surface of pulsars and accelerated by strong electric
fields to high energies if pulsars are strange stars with the
crusts, comprised of nuggets embedded in a uniform electron
background. Such high energy nuggets called usually strangelets
give an observable contribution into galactic cosmic rays and may
be detected by the upcoming cosmic ray experiment Alpha Magnetic
Spectrometer AMS-02 on the International Space Station.
\end{abstract}

% PACS, the Physics and Astronomy
                             % Classification Scheme.
%\keywords{Suggested keywords}%Use showkeys class option if keyword
                              %display desired
\maketitle

\section{\label{sec:level1}INTRODUCTION}
It has long been suggested that strange quark matter (SQM)
composed of roughly equal numbers of up, down, and strange quarks
plus a small number of electrons (to neutralize the electric
charge of the quarks) may be absolutely stable, i.e., stronger
bound than iron \cite{B71,W84}. Later, predictions of increased
SQM stability in some color-superconducting phases have made this
conjecture more likely than previously believed \cite{ABR01}. For
lumps of SQM the possible range of baryon number $A$ is from $\sim
10^2-10^3$ (strangelets) to a few times $10^{57}$ (strange stars).
Strangelets may exist in at least three possible varieties:
"ordinary" (unpaired) strangelets \cite{W84}, color-flavor locked
strangelets \cite{M01}, and two-flavor paired strangelets
\cite{AR06}. Strangelets may be accelerated to very high energies
similar to atomic nuclei and observed as a component of cosmic
rays (for review, see \cite{M05}). Searching for strangelets in
cosmic rays is one of the best ways to test the possible stability
of SQM.

At present, the production of strangelets and their subsequent
acceleration are poorly known. It is usually assumed that
strangelets are an outcome of strange star collisions \cite{M88}
and the Type II supernova explosions, where strange stars form
\cite{BH89}. No realistic simulations of such processes have been
performed to date. The value of the galactic production rate of
strangelets used in different calculations of the strangelet flux
in cosmic rays is rather an assumption than a calculable result
and differ by several orders of magnitude \cite{M05,BH89,M05b}.
Fermi acceleration in supernova shocks is considered as the
mechanism of strangelet acceleration \cite{BH89,M05b}. However,
for Fermi acceleration there is a well known problem of particle
ejection. The point is that only supra-thermal particles may be
accelerated by shocks, and therefore, for acceleration by shocks
the initial energy of strangelets has to be high enough. Since the
energy distribution of strangelets formed in collisions of strange
stars and supernova explosions is unknown, we have additional
uncertainty in predictions of strangelet contribution in cosmic
rays \cite{addM}. In this paper we consider other sources
(pulsars) of high energy strangelets in the Galaxy. We estimate
the mean production rate of strangelets and their typical energies
within a factor of 2-3 or so.

\section{\label{sec:level2}STRANGELET PRODUCTION}

Following \cite{M05,M05b} we assume that SQM is the ground state
of strong interaction, and all compact objects identified with
neutron stars (including radio pulsars) are, in fact, strange
stars. Recently, it was shown that if Debye screening and surface
tension may be neglected then the SQM surface of a strange star
must actually fragment into a charge-separated mixture, involving
positively-charged strangelets immersed in a negatively-charged
background of electrons \cite{JRS06,AR06}. This results in
formation of a normal-matter-like crust where strangelets are
instead of atomic nuclei. For strangelets in such a crust the
expected values of $A$ and $Z$ are $\sim 10^3$ and $\sim 10^2$,
respectively (cf. \S~4).

Strong electric fields are generated in the magnetosphere of a
radio pulsar because of its rotation (e.g., \cite{M91}). The sign
of the charge of particles that tend to be ejected from the pulsar
surface by this field depends on the sign of ${\bf \Omega \cdot
B}$, where ${\bf\Omega}$ is the angular velocity of the pulsar
rotation and ${\bf B}$ is the surface magnetic field. If ${\bf
\Omega \cdot B}<0$ positively charged particles (in our case,
strangelets) are ejected, and the rate of strangelet ejection from
the pulsar crust is
\begin{equation}
\dot N_{\rm str}\simeq 2\pi r_p^2n_{\rm cr}c\simeq
{\Omega^2R^3B_p\cos \chi\over cZe} \,, \label{dotN}
\end{equation}
where $r_p\simeq (\Omega R/c)^{1/2}R$ is the radius of the polar
cap around the magnetic pole from where strangelets flow away,
$\Omega =|{\bf \Omega}|$, $R\simeq 10^6$ cm is the radius of the
strange star, $B_p$ is the magnetic field strength at the magnetic
pole, $\chi$ is the angle between the rotational and magnetic
axis, and
\begin{equation}
n_{\rm cr}\simeq {{\bf \Omega \cdot B}_p\over 2\pi cZe}\simeq {
\Omega  B_p\cos\chi\over 2\pi cZe} \label{ncr}
\end{equation}
is the strangelet density that is necessary for screening of the
electric field component along the magnetic field in the vicinity
of the magnetic pole \cite{RS75}. In spite of that the estimate of
$\dot N_{\rm str}$ is written here for a dipole magnetic field, it
is also valid for more realistic magnetic fields. This is because
the combination $\pi r_p^2B_p$, which gives the total magnetic
flux penetrating the pulsar light cylinder, is the value measured
from deceleration of the pulsar rotation and does not depend on
how the magnetic field varies from the light cylinder to the
pulsar surface \cite{RS75}.

The total number of strangelets ejected from a pulsar from the
time ($t_0=0$) of its formation at angular velocity $\Omega_0$ to
the time $t$ when the angular velocity is $\Omega$ may be written
as
\begin{equation}
N_{\rm str}=\int_0^t\dot N_{\rm str}\,dt=\int_{\Omega_0}^\Omega
{\dot N_{\rm str}\over \dot \Omega}\,d\Omega\,.\label{Nstr}
\end{equation}
where
\begin{equation}
\dot \Omega\simeq -{\Omega^3R^6B_p^2\over c^3I} \label{dOt}
\end{equation}
is the rate of the angular velocity decrease (e.g.,
\cite{M91,RS75,S06}), and $I$ is the moment of inertia of the
strange star.

Eqs. (\ref{dotN}), (\ref{Nstr}) and (\ref{dOt}) yield
\begin{equation}
N_{\rm str}\simeq {c^2I\cos\chi\over Ze R^3B_p} \ln {\Omega_0\over
\Omega}\,. \label{totalNstr}
\end{equation}
The value of $N_{\rm str}$ weakly depends on the angular velocity
that varies from $\Omega_0\simeq 10^3-10^4$ s$^{-1}$ at the pulsar
formation to $\Omega_f\simeq 10^{-1}$ s$^{-1}$ when the pulsar age
is comparable with the age of the Universe ($\sim 10^{10}$ yr).

Substituting $\ln (\Omega_0/\Omega_f)\simeq 10$ into Eq.
(\ref{totalNstr}) we have the following estimate for the total
number of strangelets ejected from a pulsar during its live:
\begin{equation}
N_{\rm str}\simeq {5c^2I\over Ze R^3B_p}\,. \label{totalNstrfinal}
\end{equation}
Here we assume that directions of the rotational and magnetic axes
of pulsars are independent, and the average value of $\cos\chi$ is
1/2.

Taking into account that the birthrate of pulsars is $\eta\simeq
1.4\times 10^{-2}$ yr$^{-1}$ \cite{L06}, the mean production rate
of strangelets in our Galaxy by pulsars is
\begin{eqnarray}
\dot {\mathfrak N}_{\rm str}\simeq \eta N_{\rm str} \simeq
1.3\times 10^{42}\Lambda\,\,\,{\rm yr}^{-1}\,. \label{totalrateG}
\end{eqnarray}
where
\begin{eqnarray}
\Lambda=\left({Z\over 10^2}\right)^{-1}\left({I\over 10^{45}\,
{\rm g~cm}^2}\right)\left({R\over 10^6\,{\rm
cm}}\right)^{-3}\left({B_p\over 10^{12}\,{\rm G}}\right)^{-1}
\label{Lambda}
\end{eqnarray}
is near unity for typical choices of parameters.

If the assumptions formulated in the beginning of this session on
the stability of SQM and the properties of strange stars are
valid, Eq. (\ref{totalrateG}) gives a low limit on the production
rate of strangelets in the Galaxy because other mechanisms
\cite{M88,BH89} may produce strangelets as well.

The production rate of mass in strangelets with $A\simeq 10^3$ is
$\dot {\mathfrak M}_{\rm str}\simeq Am_p\dot {\mathfrak N}\simeq
10^{-12}\Lambda\,M_\odot\,{\rm yr}^{-1}$ that is hundred times
smaller than the value suggested by Madsen \cite{M05,M05b}, where
$m_p$ is the proton mass. The value of ${\mathfrak M}_{\rm str}$
increases with increase of $A$ because the ratio $A/Z$ increases.
For "ordinary" strangelets ($Z\simeq 8A^{1/3}$) with $A\simeq
10^6$ (if they exist in the crusts) we have $\dot {\mathfrak
M}_{\rm str}\simeq 10^{-10}\,M_\odot\,{\rm yr}^{-1}$ as Madsen
suggested.

\section{\label{sec:level3}STRANGELET ACCELERATION}

It is now commonly accepted that in the pulsar magnetospheres
there are two regions where electric fields are very strong and
may accelerate outflowing particles to relativistic energies
\cite{M91}. They are located near the polar caps and the light
cylinders and called polar gaps and outer gaps, respectively. The
electric potential at polar gaps weakly depends on the pulsar
parameters, and it is $\Delta\Phi_p\simeq 3\times 10^{12}$~V
within a factor of 2-3 \cite{RS75,HM98}. The typical energy of
strangelets accelerated in polar gaps is
\begin{equation}
E_{\rm str}^p\simeq Ze\Delta\Phi_p\simeq 3\times
10^{14}(Z/10^2)\,\,{\rm eV}\,. \label{Ep}
\end{equation}
The injection rate of strangelets with this energy into the Galaxy
is given by Eq. (\ref{totalrateG}).

Outer gaps can operate only in the magnetospheres of young pulsars
with the period $P=2\pi/\Omega\lesssim P_d$, where $P_d$ is
somewhere between 0.1~s and 0.3~s \cite{CHR86,ZC97}. For pulsars
with $P\simeq P_d$ the polar gap size $(l_{out}^\perp)$ across the
magnetic field lines is about the radius of the light cylinder,
i.e., $\sim c/\Omega$. In this case the main part of strangelets
outflowing from the polar cap may be accelerated in the outer gap.
At $P<P_d$ the value of $l_{out}^\perp$ decreases with decrease of
$P$ ($l_{out}^\perp \propto P^\alpha$, where $\alpha$ is $\sim
1.2-1.3$ \cite{ZC97}), and the fraction of strangelets accelerated
in the outer gap deceases with decrease of $P$ too. As a result,
in the process of deceleration of the pulsar rotation the outer
gap accelerates strangelets mainly at $P\sim P_d$ with more or
less the same efficiency as the polar gap. Therefore, the total
number of strangelets accelerated in the outer gap of a pulsar is
nearly the value given by Eq. (\ref{totalNstr}) only without $\ln
(\Omega_0/\Omega)$ that is $\sim 10$. Hence, in the Galaxy the
mean production rate of strangelets accelerated in outer gaps of
pulsars is $\sim 0.1\,\dot {\mathfrak N}_{\rm str}$.

The typical energy of strangelets accelerated in outer gaps is
\begin{equation}
E_{\rm str}^{out}\simeq Ze\Delta\Phi_{out}\simeq 3\times 10^{16}
(Z/10^2)\,\,{\rm eV}\,. \label{Eouter}
\end{equation}
where $\Delta\Phi_{out}\simeq 3\times 10^{14}$~V is the outer gap
potential for a pulsar at $P\simeq P_d$ \cite{CHR86,ZC97}.

\section{\label{sec:level3}STRANGELETS IN COSMIC RAYS AND THEIR DETECTION}

Strangelets ejected from pulsars undergo many processes such
spallation, energy losses, and escape from the Galaxy. In our case
when the strangelet energies given by Eqs. (\ref{Ep}) and
(\ref{Eouter}) are significantly higher than $10^{11}Z\,$~eV, the
escape process dominates over others and determines the density of
strangelets in the Galaxy \cite{M05b}.

At energies of $\sim 10^{14}-10^{15}$ eV the expected flux of
strangelets accelerated in the polar gaps of pulsars is
\begin{eqnarray}
F_{\rm str}^p\simeq {\dot {\mathfrak N}_{\rm str}\tau_{\rm
esc}(E_{\rm str}^p,Z)\,c\over 4\pi V }%\nonumber\\
\simeq 25 \,\Lambda\,\,\,({\rm m}^{2}\,{\rm sterad}~{\rm
yr})^{-1}, \label{fluxEp}
\end{eqnarray}
where
\begin{equation}
\tau_{esc}(E,Z)\simeq {1.7\times 10^5\over n}\left({E\over E_{\rm
str}^p}\right)^{-0.6}\,\,{\rm yr} \label{tauEsc}
\end{equation}
is the average escape time from the Galaxy for particles with the
energy $E$ and the charge $Z$ \cite{M05b}, $V$ is the galactic
volume, and $n$ is the average hydrogen number density of the
interstellar medium per cm$^3$. To obtain the estimate Eq.
(\ref{fluxEp}) we have used that the mass of interstellar hydrogen
in the Galaxy is $m_pnV\simeq 5.5\times 10^9\,M_\odot$ \cite{C00}.

At high energies ($\sim 10^{16}-10^{17}$ eV) the expected flux of
strangelets accelerated in the outer gaps of pulsars is
\begin{eqnarray}
F_{\rm str}^{out}\simeq 0.1\,\Lambda\,\,\,{\rm m}^{-2}\,{\rm
yr}^{-1}\,{\rm sterad}^{-1} \,, \label{fluxEout}
\end{eqnarray}

The parameters ($Z\simeq 10^2$ and $A\simeq 10^3$) of strangelets
ejected from the crusts of strange stars (pulsars) and accelerated
in polar gaps are within the interesting region for the upcoming
cosmic ray experiment Alpha Magnetic Spectrometer AMS-02 on the
International Space Station \cite{S04}. The acceptance of AMP-02
is about 0.5 m$^2$ sterad. AMS-02 will analyze the flux of cosmic
rays in unprecedented details for three years. The strangelet flux
(\ref{fluxEp}) is high enough to be detected by AMS-02, especially
if the mass ($m_s$) of strange quarks is rather small (cf.
\cite{ms}). The point is that for strangelets the ratio $Z/A$ is
proportional to $m_s^2$ \cite{M05c}. We used the strangelet
parameters calculated for $m_s=200$~MeV \cite{AR06}. However,
$m_s$ may differ from this value by a factor of 2-3 or so, and $Z$
may be at least several times smaller than the value we used. This
may simplify detection of strangelets by AMS-02 because the flux
$F_{\rm str}^p$ increases with decrease of $Z$ while the energy of
strangelets decreases. Besides, for the particle charge it is easy
to be measured by AMS-02 if it is not too high (see \cite{S04}).

Plausibly, the polar gaps of pulsars are non-stationary
\cite{RS75}, and an essential part of the out-flowing strangelets
have energies that are significantly smaller than the value given
by Eq. (\ref{Ep}). This favors detection of strangelets ejected
from strange stars (pulsars) by AMS-02.

The flux of high energy strangelets (\ref{fluxEout}) is of the
order of the flux of cosmic rays at the same energy \cite{H03} and
may be detected in future.

\begin{acknowledgments}
V.V.U. thanks the Department of Physics, University of Hong Kong,
where this work in part was carried out, for its kind hospitality.
This work was supported by the Israel Science Foundation of the
Israel Academy of Sciences and Humanities and a RGC grant of Hong
Kong Government under HKU 7013/06P.
\end{acknowledgments}

%\bibliography{apssamp}% Produces the bibliography via BibTeX.

\begin{references}
\bibitem{addM} For strangelets the ratio
$A/Z$ is several times larger that the same for nuclei, and
therefore, strangelets are more efficiently injected into an
accelerating shock than nuclei \cite{M05b}. However, for
acceleration of strangelets by supernova shocks the injection
problem remains.
\bibitem{ms} The strange quark mass $m_s$ cannot be too small.
Otherwise, the mixed phase does not form at the surface of strange
stars \cite{JRS06}.
\end{references}
%%%%%%%%%%%%%%%%%%%%%%%%%%%%%%%%%%%%%%%%%%%%%%%%%%%%%%%%%%%%%%%%%%%%%

%%%%%%%%%%%%%%%%%%%%%%%%%%%%%%%%%%%%%%%%%%%%%%%%%%%%%%%%%%%%%%%%%%%%%

%%%%%%%%%%%%%%%%%%%%%%%%%%%%%%%%%%%%%%%%%%%%%%%%%%%%%%%%%%%%%%%%%%%%
\end{document}